\documentclass[twocolumn]{aastex631}
\usepackage{amsmath}
\usepackage{bbold}
\usepackage{hyperref}
\usepackage{lipsum}
\usepackage{blindtext}
\usepackage{multirow}
\usepackage[dvipsnames]{xcolor}

\usepackage{mathtools}

\newcommand{\lz}{{low-redshift}}

\begin{document}

\title{Unveiling dark energy properties with high-sensitivity cross-correlations of neutral hydrogen intensity mapping and galaxy surveys}

\author[0000-0002-6519-6038]{Alessandro Marins}
\affiliation{Department of Astronomy,
University of Science and Technology of China, Hefei 230026, China}
\affiliation{School of Astronomy and Space Science, University of Science and Technology of China, Hefei 230026, China}

\author[0000-0001-7438-5896]{Chang Feng}
\altaffiliation{Corresponding author: changfeng@ustc.edu.cn}
\affiliation{Department of Astronomy,
University of Science and Technology of China, Hefei 230026, China}
\affiliation{School of Astronomy and Space Science, University of Science and Technology of China, Hefei 230026, China}

\author[0000-0003-2063-4345]{Filipe B. Abdalla}
\affiliation{Department of Astronomy,
University of Science and Technology of China, Hefei 230026, China}
\affiliation{School of Astronomy and Space Science, University of Science and Technology of China, Hefei 230026, China}

\begin{abstract}
The redshifted 21 cm line from the hyperfine structure of neutral hydrogen atoms is a promising tracer for the three-dimensional evolution of our universe. Its broad spatial and temporal coverage is crucial for understanding the complex nature of dark matter and dark energy. However, it is very challenging to directly detect the 21 cm signal due to the existence of radio foreground contaminants that are orders of magnitude brighter. Therefore, mitigating the foreground contamination becomes an indispensable task for detecting the 21 cm signal, which is also expected to be correlated with the dark-matter-dominated large-scale structure. The cross-correlations between the neutral hydrogen intensity mapping and galaxy surveys in the future can not only confirm a detection of the 21 cm signal but can also be a complementary probe for 21 cm cosmology. To meet the precision requirements for cosmological studies, it is important to investigate the complex features of the estimated cross-correlations using numerical simulations. In this work, we simulate the low-redshift HI observations and narrow-band optical surveys in the future and obtain the foreground- and bias-mitigated HI-galaxy cross-power spectra, with which we perform a Bayesian analysis to infer cosmological parameters of dynamical dark energy model. The method developed in this work will be important for cosmological studies with the 21 cm intensity mapping. 
\end{abstract}

\section{Introduction}

Neutral hydrogen (HI) atoms formed when the free electrons of hot plasma in the early Universe were captured by the protons, and this recombination process also released the photons that were coupled to the free electrons, forming the cosmic microwave background (CMB). Since then, the neutral hydrogen started to illuminate the entire Universe with electromagnetic radiation at a unique 21 cm wavelength arising from the hyperfine structure. This rest-frame wavelength has been stretched as the Universe is expanding; thus, the redshifted 21 cm signals can trace different epochs of the Universe within a broad temporal and spatial space, serving as a powerful cosmological probe to many fundamental physics problems~\citep{21cmcosmo1,21cmcosmo2,21cmcosmo3}. However, the redshifted 21 cm signals are extremely faint and highly contaminated by both the Galactic and extragalactic radio emissions. Therefore, an optimized strategy with precise foreground and systematics mitigations is required to not only make a detection but also use the measured 21 cm signals to perform cosmological studies. 

So far, the series of internal linear combination (ILC) has been widely adopted for foreground removal tasks in CMB data, assuming the foreground contaminants have smooth spectral energy distributions (SEDs)~\citep{ilc}. Extensions of the ILC method, such as needlet ILC~\citep[NILC;][]{nilc} and generalized needlet ILC~\citep[GNILC;][]{gnilc}, as well as other blind algorithms such as principal component analysis~\citep[PCA;][]{chang2010} and independent component analysis~\citep[ICA;][]{maino2002}, have been developed and adapted to the 21 cm signal~\citep{chapman2012, olivari2016, fast_2021_yohana}. The essence of these approaches is to form linear combinations of frequency maps whose weights are chosen to minimize the variance of the observational maps while preserving the target signal. The result of this procedure is the estimation of foreground emissions that are then subtracted from the observations, leaving the 21 cm emission as part of the residual, together with instrumental noise and other unmodeled components. These algorithms have passed many tests and have proven to effectively detect the 21 cm signals using the auto correlations from multifrequency data~\citep{marins2022}.

However, measurements of auto correlations are challenged by several issues, such as the requirement for long observing time to reach sufficiently high sensitivities, the debiasing of the instrumental systematics, and the mitigation of radio-frequency interference. On the other hand, these methods also face other challenges when the reconstructed 21 cm signals are cross-correlated with external datasets, as found in recent studies~\citep{marins2025, signalosstheory}; the linear combinations of different frequency maps can inevitably suppress the long-wavelength radial modes of the HI fluctuations, causing significant signal loss. 

Cross correlation has been adopted as an effective approach to detecting the faint 21 cm signals. The evidence of 21 cm signals was found by cross-correlating the HI observations after some integration time with the optical surveys, which have better sensitivities to trace the large-scale structure (LSS)~\citep{chang2010, masui2013, anderson2018, wolz2022, cunnington2023}. In addition to making a detection of the 21 cm signals by cross correlations, we should also investigate the inference of the cosmological parameters using the reconstructed 21 cm signals after complex procedures of foreground- and bias-mitigation. 

In previous work (\cite{marins2025}; hereafter referred to as AM2025), we cross-correlated the foreground-removed HI with the weak lensing maps and found different levels of signal loss in different scenarios. By making numerical tests and breaking down the signal loss, we found that a broad redshift coverage of the convergence field is the main reason for such a significant signal loss since it makes the convergence field perfectly correlated across all HI channels and can induce HI-signal leakage after performing the foreground removal. In light of this, we can reduce the signal loss by choosing the LSS tracers observed within narrow redshift ranges, such as the narrow-band galaxy surveys. Thereby, we can obtain precise measurements of the cross correlations, which will be invaluable for studying dark energy.

Dark energy is believed to accelerate the expansion of the late Universe; however, its nature remains unknown. Especially, the recent measurements of the Dark Energy Spectroscopic Instrument (DESI) data release 2 (DR2) revealed intriguing evidence of a dynamical dark energy with a phantom-cross around redshift $z\sim0.4$, making the understanding of dark energy more complicated~\citep{desi2}. The cross-correlations between HI and LSS tracers can provide independent measurements of structure growth, thus, are important for testing dark energy theories.

In this work, we for the first time test the inference of the cosmological parameters for the dynamical dark energy models using the foreground- and bias-mitigated HI-galaxy cross-power spectra. The work is organized as follows. We introduce the theoretical models in Sec. \ref{model}, and experiment settings in Sec. \ref{experiment}. We discuss the results of numerical tests in Sec. \ref{validations} and make forecasts for testing the dynamical dark energy models with future cross-power spectra in Sec. \ref{forecast}. We conclude in Sec. \ref{conclu}.

\section{Theoretical calculations of the cross-correlations between neutral hydrogen intensity mapping and narrow-band galaxy surveys}
\label{model}
The dark-matter dominated large-scale structure can be traced by both radio and optical surveys. In this work, we consider sky patches covered by both a low-redshift intensity mapping (IM) experiment and an optical survey in the future, and study the cross-correlation signals between the two measurements.

We assume that the HI brightness temperature is a tracer of the underlying dark matter distribution, i.e.,
\begin{equation}
    \delta T_b ({\bf x})=b_{\rm HI} \delta({\bf x}),
\end{equation} where $b_{\rm HI}$ is the HI bias and $\bf x$ denotes the real-space coordinates. To simplify the theoretical modeling of the HI signals, we do not consider the redshift-space distortions and intrinsic alignment. Also, we assume the HI bias factor $b_{\rm HI}$ is a constant and does not evolve over time. For the IM experiment, each frequency channel measures the fluctuations within a narrow redshift range, and the corresponding redshift kernel is 
\begin{eqnarray}
    W_{\textrm{HI},j}\!\left(z\right) &=& b_{\textrm{HI}}(z) \bar{T}_{\textrm{HI}}(z)W_{\Delta_j}\!(z),
    \label{eq: HI kernel}
\end{eqnarray} with $\bar T_{\rm HI}=188.8hH_0/H(z)\Omega_{\rm HI}(1+z)^2\rm mK$, $\Omega_{\textrm{HI}}(z)=4.86\times10^{-4}$, and $b_{\textrm{HI}}(z)=1$~\citep{phibull}. Here, $h=H_0/100$, $H_0$ is the Hubble constant, and $H(z)$ is the Hubble parameter. $W_{\Delta_j} (z)$ is the window function derived from the radio bandwidth $\Delta_{\tiny j}$, which is assumed to be a top-hat window.

\begin{figure}
    \centering
    \includegraphics[scale=0.29]{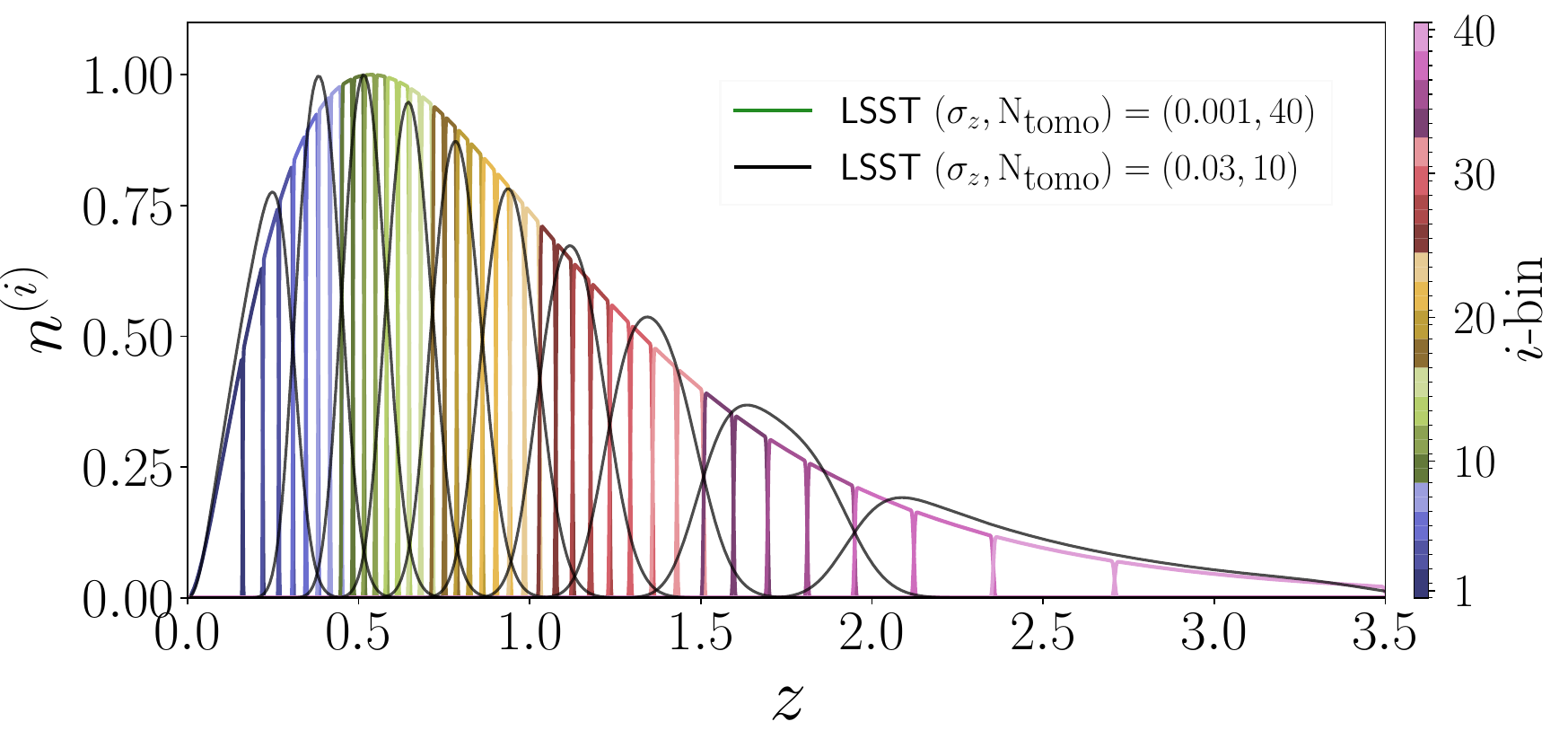}
    \caption{Photometric redshift distributions of an LSST-like galaxy survey assuming two different combinations of photometric error and galaxy binning.}
    \label{fig: dNdz_i}
\end{figure}

We consider an optical survey with narrow bands to detect galaxy samples, from which the galaxy distribution is assumed to be $n(z)\sim z^{\beta}e^{-(z/z_0)^\alpha}$, where $(z_0, \alpha,\beta)$ are free parameters~\citep{bartelmann2001}. For each band, the galaxy density distribution is normalized to a unit area in the $i$th redshift bin bounded by the values $z^i_{\min}$ and $z^i_{\max}$~\citep{Ma2006}, i.e., 
\begin{eqnarray}
    n^{(i)}(z) \sim\int_{z^i_{\min}}^{z^i_{\max}}\!\textrm{d}z_p p(z_p|z)n(z),
    \label{eq: dNdz_i}
\end{eqnarray}
where $p(z_p|z)$ is the probability of a galaxy at $z$ to be measured at $z_p$ and is expressed as 
\begin{eqnarray}
    p(z_p|z) = \frac{1}{\sqrt{2\pi}\sigma_z(z)}\exp\left[-\frac{1}{2}\left(\frac{z-z_p}{\sigma_z(z)}\right)^2\right].
    \label{eq: prob LSST}
\end{eqnarray}
Here, $\sigma_z(z)=\sigma^0_z(1+z)$ is the photometric redshift error. We assume an optical survey like the Rubin Legacy Survey of Space and Time (LSST) with a ten-year observation time, and take the same parameters as described in Table \ref{tab: LSST config}~\citep{LSST2018_science_collaboration}. We have sliced the galaxy distribution into 40 bins with the width of each bin higher than the photometric redshift error, and each band is shown in Figure \ref{fig: dNdz_i}. The minimum and maximum redshifts of each $i$-bin are constructed as $z^i_{\max}=z^i_{\min}+\Delta z$.
The galaxy kernel for each narrow band is 
\begin{equation}
W^{(i)}_{\textrm{G}}(z)=b_{\textrm{G}}(z)n^{(i)}(z),
    \label{eq: G kernel}
\end{equation}
where $b_{\textrm{G}}(z)=\sqrt{1+z}$ is the linear galaxy bias~\citep{EUCLID_blanchard2020}.

We adopt the Limber approximation~\citep{limber} to calculate the cross correlations for two generic LSS tracers $A$ and $B$ measured at two frequencies $\nu_i$ and $\nu_j$. The angular cross-power spectrum can be expressed as
\begin{equation}
C^{A, B}_{\ell}(\nu_i,\nu_j) = \int \frac{\textrm{d}\chi}{\chi^2}W^{(i)}_{A} (\chi) W^{(j)}_{B}(\chi)P\left( k, \chi\right),
\label{eq: LIMBER_FIELDS_A_B}
\end{equation}
where fluctuation wavenumbers $k$ are related to power-spectrum multipole $\ell$ and the comoving distance $\chi$ as $k=(\ell + 1/2)/\chi$, and $W^{(i)}(\chi)=W^{(i)}(z)dz/d\chi$. In this work, $A,B \in \{\textrm{HI}, g\}$. The cosmological parameter set is ($H_0$, $\Omega_{b}h^2$, $\Omega_{c}h^2$, $n_s$, $w_0$, $w_a$) = (67.5 {\rm km/s/Mpc}, 0.022, 0.122, 0.965, -1, 0). The matter power spectrum is calculated via the public code \texttt{CAMB}\footnote{https://camb.readthedocs.io/en/latest/} using the halofit model for non-linear corrections based on Takahashi's model \citep{takahashi_camb2012}. We also adopted the Parameterized Post-Friedmann Framework referred to as DarkEnergyPPF, which makes the dark energy equation of state time-dependent with two free parameters $w_0$ and $w_a$~\citep{DarkEnergyPPF_fang2008}. Although it is not a physical model, it serves as an effective approximation for any model in which the energy and momentum of dark energy are conserved independently. 
{\setlength{\tabcolsep}{10pt}
\setlength{\belowdeluxetableskip}{-24pt}
\setlength{\intextsep}{4pt}
\begin{deluxetable}{@{}ccccccc@{}}
\tablecaption{LSST parameters for the photometric redshift distribution described in Eqs. (\ref{eq: dNdz_i}) and (\ref{eq: prob LSST}). The redshifts $z_{\min}$ and $z_{\max}$ refer to the minimum and maximum redshifts covered by LSST, respectively.\label{tab: LSST config}}
\tablecolumns{7}
\tablehead{
\colhead{$z_0$} &
\colhead{$\alpha$} & \colhead{$\beta$} & \colhead{$\sigma^0_z$} &
\colhead{$\Delta z$} & \colhead{$z_{\min}$} & \colhead{$z_{\max}$}
}
\startdata
 0.28 & 0.90 & 2 & 0.001 & 0.1 & 0.001 & 3.5\\
\enddata
\end{deluxetable}
}

In this work, we assume a full-sky coverage and a range of frequencies compatible with a \lz\, IM experiment. The cross-power spectra are binned by \texttt{NaMaster} \citep{namaster} within $30<\ell<400$ at a spacing $\Delta\ell=4$, resulting in 93 bands per cross-correlation. We impose two selection criteria to exclude the cross-power spectra with low signal-to-noise ratios (SNRs): (1) $\ell(\ell+1)/2\pi C_\ell^{{\rm HI},g}(\nu_i,\nu_j)>10^{-4}$ for at least half of the multipoles and (2) $\sqrt{\sum_\ell({C}_{\ell}^{\textrm{HI},\textrm{g}}(\nu)/\Delta {C}_{\ell}^{\textrm{HI},\textrm{g}}(\nu))^2}/\mathcal{N}_{\ell}\gtrsim \epsilon$ where $\epsilon=0.6$ and $\epsilon=1$ for the cross-power spectra without and with the signal-loss corrections, respectively. Here, $\mathcal{N}_{\ell}$ denotes the number of multipoles for a cross-power spectrum. Our selection criteria exclude low SNR cross-power spectra and also remove HI channels that are affected by edge effects after the foreground removal procedure.

\begin{figure}
    \centering
    \includegraphics[scale=0.34]{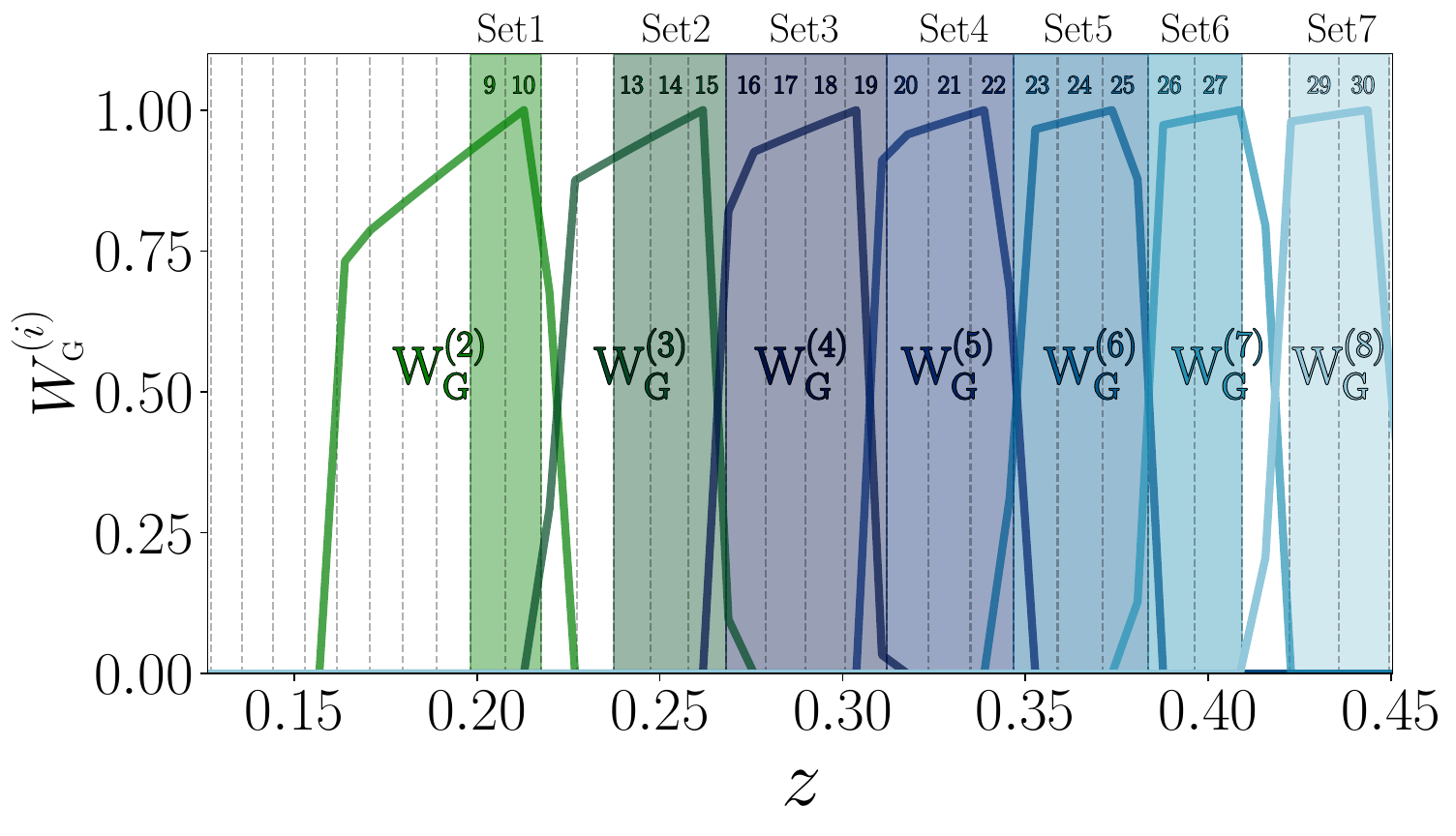}  
    \caption{HI and galaxy window functions defined in Eq. (\ref{eq: G kernel}). The vertical strips with numbers denote the HI bands, and the seven top-hat-like windows represent the narrow bands of the galaxy survey.}

    \label{fig: Wg_i}
\end{figure}

We have adopted $\ell_{\min}=30$ to avoid higher foreground leakage and to ensure that the white noise is only a subdominant component even for a short observing time~\citep{marins2022, marins2025}. The choice of $\ell_{\max}=400$ is made because of a few factors, including the beam smoothing, narrow redshift bins ($\Delta z\sim0.01$), and explicit modeling of mildly nonlinear scales, following a similar reasoning in~\cite{alonso2015ultra-large-scale-e44}. For the \lz\, IM experiment, the finite angular beam resolution progressively suppresses small-scale HI fluctuations toward high multipoles, with a significant damping around $\ell\sim300$, thereby dramatically reducing the nonlinear effects. For the galaxy survey, the tomographic bands yield sufficiently narrow and localized kernels such that $\ell=400$ corresponds to $k\sim0.3\,{\rm Mpc}^{-1}$ at $z\sim0.45$ where nonlinear effects are small. Instrumental and shot-noise contributions are also included for both surveys. In Figure \ref{fig: APS_2}, we display the theoretical cross-power spectra for two different redshifts in black dashed lines for reference. 

For the LSST-like galaxy survey, we only consider the seven narrow tomographic bands within $0.1<z<0.5$ as shown in Figure \ref{fig: Wg_i} because only these bands can pass the selection criteria discussed above. The vertical strips refer to the HI channels overlapping with the narrow bands of the galaxy surveys. The color-coded strips thus refer to the cross-correlations between the HI channels and galaxy bands, and each of them is labeled as {\bf Set}$i$ with $i\in(1,7)$. For example, {\bf Set}$1$ corresponds to the cross-correlations between galaxy band two and the HI channels 9 and 10, spanning redshifts 0.198-0.217.

\section{Experimental configurations of both the HI IM and galaxy surveys}
\label{experiment}
We simulate the sky maps with \texttt{HEALPix} at a resolution NSIDE=256~\citep{HEALpix} and assume a white noise model for the HI IM power spectrum, i.e, $\mathcal{N}_{\ell}=\Omega_{\textrm{pix}}\sigma^2_{\textrm{pix}}$ where $\Omega_{\rm{pix}}$ is the pixel area, and $\sigma_{\textrm{pix}}$ is the noise variance per pixel which is determined by
\begin{eqnarray}
    \sigma_{\textrm{pix}} = \textrm{K}\frac{T_{\textrm{sys}}}{\sqrt{\Delta\nu t_{\textrm{pix}}}}.
    \label{eq: HI white noise}
\end{eqnarray}
Here, the noise performance is $K$, $T_{\textrm{sys}}$ is the system temperature, and $\Delta\nu$ is the bandwidth. The pixel integration time is 
\begin{eqnarray}
    t_{\textrm{pix}}= \epsilon t_{\textrm{sur}}N_{\textrm{beams}}\frac{\Omega_{\textrm{pix}}}{\Omega_{\textrm{sur}}},
\end{eqnarray}
where $\Omega_{\textrm{sur}}$ is the survey coverage area, $t_{\rm sur}$ is the total coverage time, $\epsilon$ is the coverage efficiency, and $N_{\textrm{beams}}$ is the number of beams. We also assume a Gaussian beam profile $b_{\ell}=\exp[-\ell(\ell+1)\theta^2_{ \textrm{FWHM}}/(16\log{2})]$ where the FWHM is $\theta_{\rm FWHM}$. White noise realizations are drawn from the noise power spectra calculated with the parameters listed in Table \ref{tab: requirement}.

\begin{table}
\footnotesize
\centering
\caption{Setup of the HI IM surveys for instrumental noise based on the sensitivity level of a \lz\, IM experiment~\citep{marins2025}.}
\begin{tabular}{l c }
\cline{1-2}
Angular resolution ($\theta_{ \textrm{\tiny FWHM}}$)  & 40'\\ 
Frequency range      & $980-1260$ MHz \\
Number of channels ($\textrm{n}_\nu$)  & $30$  \\
Bandwidth ($\Delta\nu$)  & $9.33$ MHz\\
Sky coverage ($\Omega_\textrm{sur}$)   & 41253 deg$^2$ \\
Number of beams ($N_{\textrm{beams}}$) & 1084 \\
System temperature ($\textrm{T}_\textrm{sys}$)  & $70$ K  \\
Total coverage time ($t_\textrm{sur}$)  & 1 year  \\
Coverage efficiency ($\epsilon$) & 1 \\
Noise performance (K) & $\sqrt{2}$  \\
\cline{1-2}
\end{tabular}
\label{tab: requirement}
\end{table}

Due to the finite sample of resolved sources, the shot noise contribution must be considered for the galaxy survey as well. It can be described as a Poisson noise with $\sigma^2_{\rm S}=1/n^{\textrm{eff}}_{\textrm{gal}, i}$, where $n^{\textrm{eff}}_{\textrm{gal}, i}$ is the effective number density of galaxies per tomographic band and is determined by $n^{\textrm{eff}}_{\textrm{gal}, i}=n_{\textrm{gal}, i}/N_{\textrm{bins}}$. Here, $n_{\textrm{gal}}$ is the surface number density of galaxies and $N_{\textrm{bins}}$ is the number of tomographic bands. For this work, we used a fixed value of the number density of galaxies for all bins of $n_{\textrm{gal}}=48\ {\rm arcmin}^{-2}$~\citep{LSST2018, LSST2018_science_collaboration}.

\section{Map simulations and foreground Removal}
\label{validations}
Due to the fact that the CMB lensing kernel overlaps broadly with the HI IM kernel along the line of sight, there is a significant signal loss in the cross correlation between HI IM and CMB lensing, as found in AM2025, where a signal loss in the cross correlation was found to be reduced when the shear kernel only partially overlaps with the HI signal. Motivated by this result, we choose narrow optical bands of the galaxy survey to investigate the signal loss problem for the cross-correlation signals.

Dividing the galaxy sample into 40 narrow bands with a photometric redshift uncertainty $\sigma_z = 0.001$ yields sharper, more localized redshift distributions. Therefore, the overlapping regions between HI channels and the galaxy bands are greatly reduced for each {\bf Set}$i$ defined in Figure \ref{fig: Wg_i}, boosting the cross correlation signals while minimizing the effective number of cross correlations. This mechanism indicates that a cross correlation with spectroscopic galaxy samples with much sharper redshift coverage would also be favorable. 

\subsection{Simulations of HI IM and galaxy overdensity maps}
We simulate correlated Gaussian HI and galaxy density maps with the covariance matrices calculated in Sec. \ref{model} following the procedures described in AM2025. We have validated that different Gaussian realizations of the HI and galaxy overdensity maps preserve the expected covariance between the two fields. We note that the covariance matrix of galaxy-galaxy internal correlations is neglected in this work for simplicity and further prove that the inclusion of those elements will not affect the cross-correlation amplitude corrections, as described in Appendix \ref{Section: cross-g and TF}. We defer the investigations with a full covariance matrix to future work.

The simulated radio sky maps, which are referred to as ${\rm X_{obs}}$, include the Gaussian HI maps ${\rm X_{HI}}$ and two major foreground components, i.e, the synchrotron and free-free emissions, collectively denoted by ${\rm X_{FG}}$. These foregrounds are simulated by \texttt{PySM3}\footnote{https://pysm3.readthedocs.io}
 \citep{thorne2017} for the HI channels considered in this work. For the free-free emission, we use a map template generated from \texttt{COMMANDER}\footnote{\href{https://github.com/Cosmoglobe/Commander}{https://github.com/Cosmoglobe/Commander}} \citep{PLANCK2016_commander}, which extracted a degree-scale map of free-free emission at 30 GHz from Planck 2015 data. The simulated synchrotron map is extrapolated from the 408 MHz synchrotron template following a power-law in frequency space~\citep{remazeilles2015}. For the map-level simulations, we do not add uncorrelated noise realizations to simplify the cross-correlation calculations but account for the noise-induced variance at the power-spectrum level.

\begin{figure*}
    \centering \includegraphics[scale=0.39]{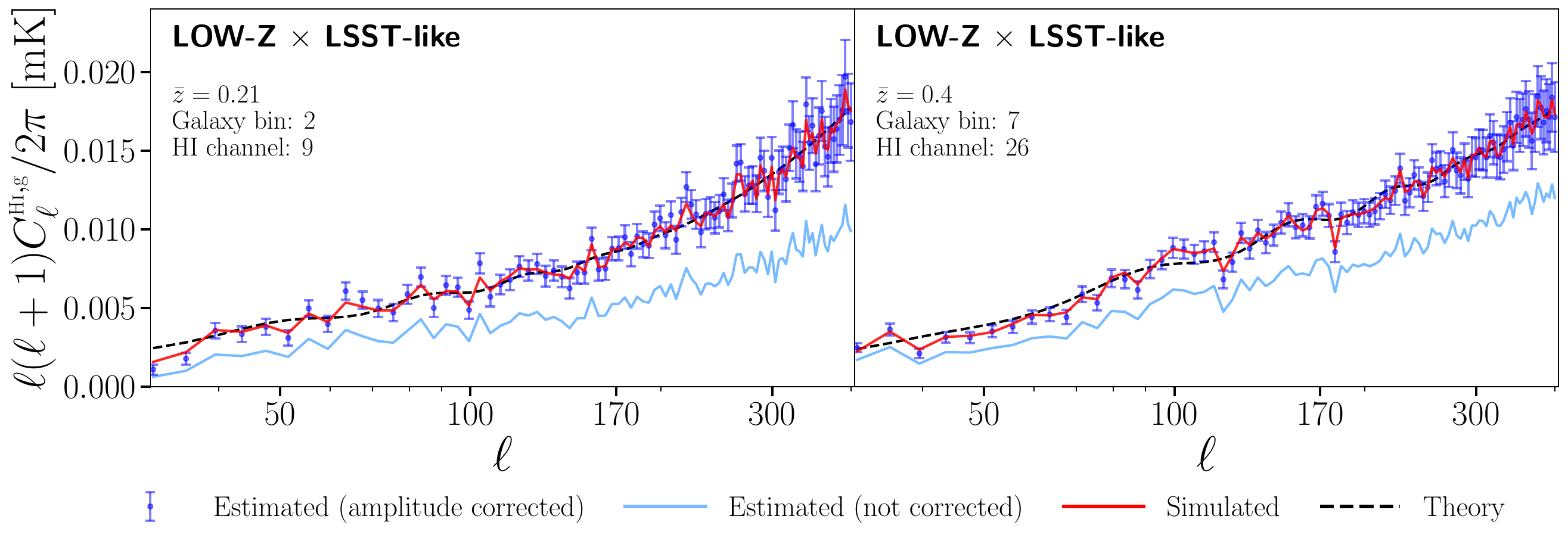}         
    \caption{Different types of cross-power spectra between HI IM and an LSST-like galaxy survey at two redshifts. The left panel corresponds to the second galaxy bin and the tenth HI channel, at a mean redshift $\bar{z}=0.21$, while the right panel corresponds to the seventh galaxy bin and the 26th HI channel, at a mean redshift  $\bar{z}=0.40$. The panels display the spectra of the model (black dashed), the signal-only Gaussian simulation (red), and the spectrum after foreground removal, with (blue) and without amplitude correction (light blue) using the transfer function (Eq. \ref{eq: tf}). The error bars are estimated from 2000 realizations.}
    \label{fig: APS_2}
\end{figure*}

\begin{figure*}
    \centering \includegraphics[scale=0.39]{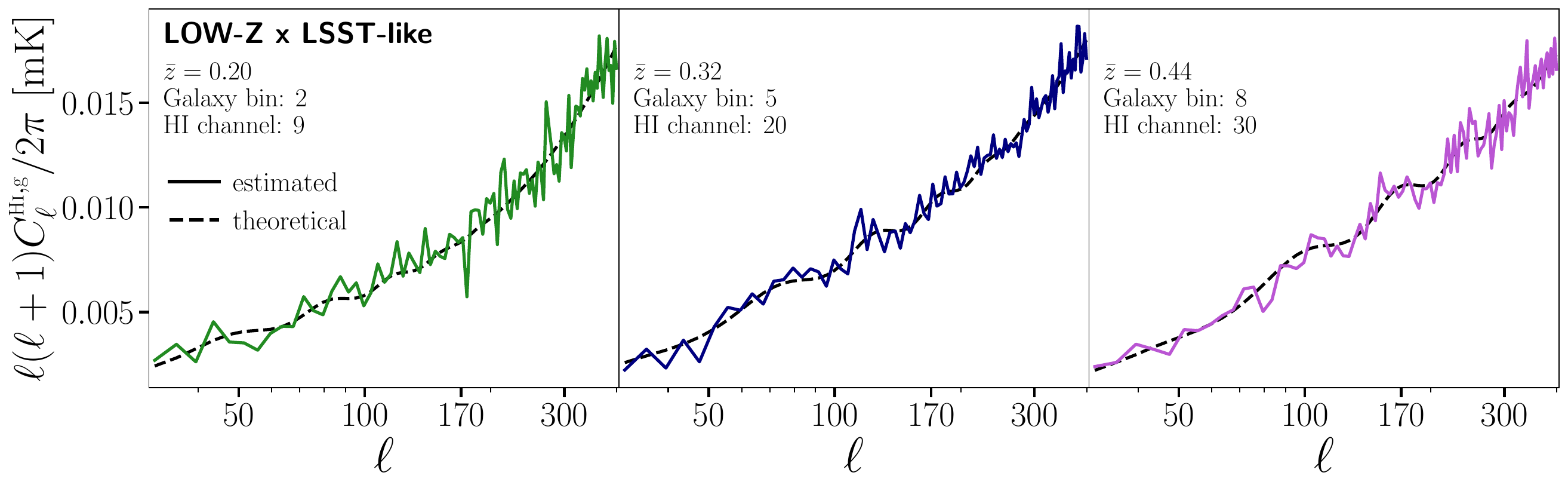}      
    \caption{Foreground- and bias-mitigated HI-galaxy cross-power spectra at multiple redshifts. The estimated cross-spectra are consistent with the theoretical calculations. }
    \label{fig: APS_1}
\end{figure*}

\subsection{Transfer functions induced by foreground-removal procedures}
As a blind method, the independent component analysis (ICA) does not rely on an explicit physical model for the foreground emission, but instead exploits only the statistical properties of the data. We adopt a fast ICA (FastICA) method to perform a computationally efficient blind foreground removal in this work. The FastICA iteratively forms an optimal linear filter that operates on the matrix of observational maps and provides an estimate of the total foreground emission. Previous studies have shown that the algorithm can recover the HI signal with satisfying accuracy~\citep{maino2002, alonso2015blind-a18, wolz2022}.

Other blind algorithms, such as generalized morphological component analysis~\citep[GMCA;][]{carucci2020}, rely on morphological diversity and sparsity of the data. The generalized needlet internal linear combination approach~\citep[GNILC;][]{olivari2016, BINGO_V, silva2025detectability-36d} estimates the effective mixing-matrix dimension by combining harmonic and spatial localizations. Moreover, parametric templates of the foregrounds have been tested with simulations, but have not achieved a similar performance as the blind algorithms~\citep{bigot-sazy2015simulations-03b, spinelli2021skao-06e}. 

The long-wavelength radial fluctuations of the HI and foreground fields are degenerate, and foreground removal procedures can remove both long-wavelength modes, causing amplitude suppression in the cross-spectra of the foreground-removed HI and the LSS tracers. To correct the amplitudes, we simulate 500 Gaussian realizations for the HI field, process these maps with the FastICA procedure, and cross-correlate the foreground-removed HI maps with the galaxy overdensity maps to obtain the cross-power spectra. Thus, the transfer functions can be estimated by comparing the simulated cross-power spectra with the input ones~\citep{switzer2013}. 

To define the transfer function calculations, let ${\rm X}$ represent the simulated sky maps, which can be described as 
\begin{equation}
    X^{\rm obs}=AS+X^{\rm HI{(0)}}
\end{equation} where the mixing matrix $A$ describes the spectral energy distributions (SEDs) of the diffuse foreground components, and $S$ describes their spatial response, and $X^{\rm HI{(0)}}$ is the input HI signal at each frequency.  We perform foreground removal using FastICA \citep{hyvarinen1999, maino2002, chapman2012}, assuming that the spatial distributions of the foregrounds are statistically independent. In this way, FastICA describes the total foreground emission through a given number of independent components by estimating a matrix $\tilde{W}$ with rows $\textbf{w}_i$ which can make the field $y_i=\textbf{w}_iX$ maximally non-Gaussian, and then estimate the mixing matrix $\hat{A}$  as the pseudo-inverse of $\tilde{W}$. In this work, we used the FastICA code via the \texttt{scikit-learn} 
Python package with four independent components (see Appendix C of \cite{marins2022}), and define the foreground-removed maps as \begin{equation}{\mathcal{F}_{\rm ICA}[\rm X]}= (1-\rm W_{FG}) {\rm X}=X^{\rm HI{(1)}},\end{equation}with the matrix~\citep{marins2025} 
\begin{eqnarray}
\rm W_{FG}=\hat{A}(\hat{A}^T\hat{A})^{-1}\hat{A}^T.    
\label{eq: W_fg}
\end{eqnarray}
We denote ${\rm X'_{HI}}$ as an independent HI mock realization introduced to quantify the HI signal loss caused by the foreground removal procedure. The basic idea is to inject this HI mock into the observational maps, perform foreground removal on the mock data with and without the injected signal, and then compute the difference between the two foreground-removed maps to estimate the lost HI modes. Specifically, we compute \begin{equation} 
X^{\rm HI{(2)}}=\mathcal{F}_{\rm ICA'}[{\rm X'}_{\rm HI}+{\rm X}^{\rm obs}]-\mathcal{F}_{\rm ICA}[{\rm X}^{\rm obs}].\end{equation} The prime on ${\rm ICA}'$ emphasizes that the two cleaning operations result in different matrices of Eq. (\ref{eq: W_fg}). 
A complementary discussion about the procedure can be found in~\cite{cunnington2026revealing-d2f}. We cross-correlate the spherical harmonic coefficients of HI and galaxy maps in each realization to get the power spectrum defined as \begin{equation}\langle\hat X^{\rm HI, \ast}_{\ell m}g_{\ell'm'}\rangle=\hat C_\ell^{{\rm HI},g}\delta_{\ell\ell'}\delta_{mm'}.
\end{equation} The transfer function is thus calculated from
\begin{eqnarray}
    \mathcal{T}_\ell(\nu) = \Bigg\langle \frac{C_\ell^{{\rm HI'},g'}(\nu)}{C_\ell^{{\rm HI}{(2)},g'}(\nu)}\Bigg\rangle,
    \label{eq: tf}
\end{eqnarray}
where the bracket denotes the ensemble average and $g^{\prime}$ refers to the galaxy simulation correlated with $X'_{\rm HI}$. The amplitude-corrected cross-power spectrum is $\hat{C}_\ell^{{\rm HI},g}(\nu) = \mathcal{T}_\ell(\nu)C_\ell^{{\rm HI}{(1)},g}(\nu)$. 

The estimation of the $A$ and $W_{\rm FG}$ matrices is primarily influenced by the level of the foreground. However, at higher redshifts covered by HI channels, edge effects can introduce small variations that are realization-dependent. In order to reduce this realization dependence in the transfer function estimation, we can average over multiple realizations to preserve the main amplitude and enhance robustness against the choice of any particular mock model, as shown in \cite{Chen2025ASurveys}, although it may also result in increased variance. To further address the high variance in the transfer function estimation, we use a Huber estimator \citep{venables2013} in Eq. (\ref{eq: tf}). This method downweights outliers based on the distribution of angular power spectra from different multipoles and frequency pairs obtained with 2000 realizations.

We show a representative plot in Figure \ref{fig: APS_2} for the cross-power spectra at $z=0.21$ and $z=0.40$. We show the cross-power spectra from theoretical calculations and signa-only Gaussian realizations in dashed black lines and red solid lines, respectively. We also show the amplitude-corrected cross-power spectra in blue and the ones with no corrections in light blue. The band-power errors are estimated from the standard deviations of 2000 cross-power spectra. The amplitude-corrected cross-power spectra are consistent with the input at different redshifts as seen from Fig. \ref{fig: APS_1}.

\begin{figure*}[htb!]
    \centering
    \includegraphics[scale=0.43]{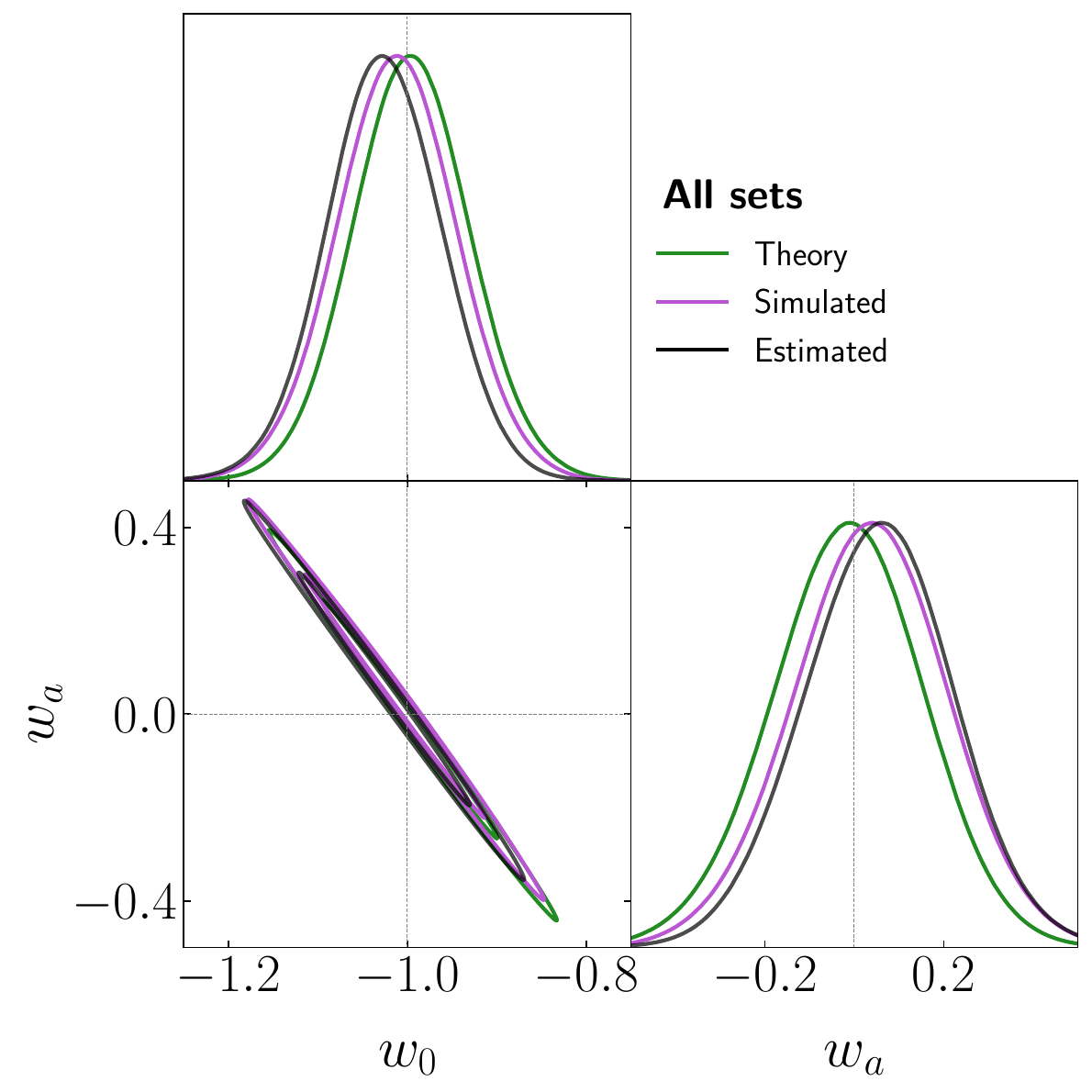} 
    \includegraphics[scale=0.43]{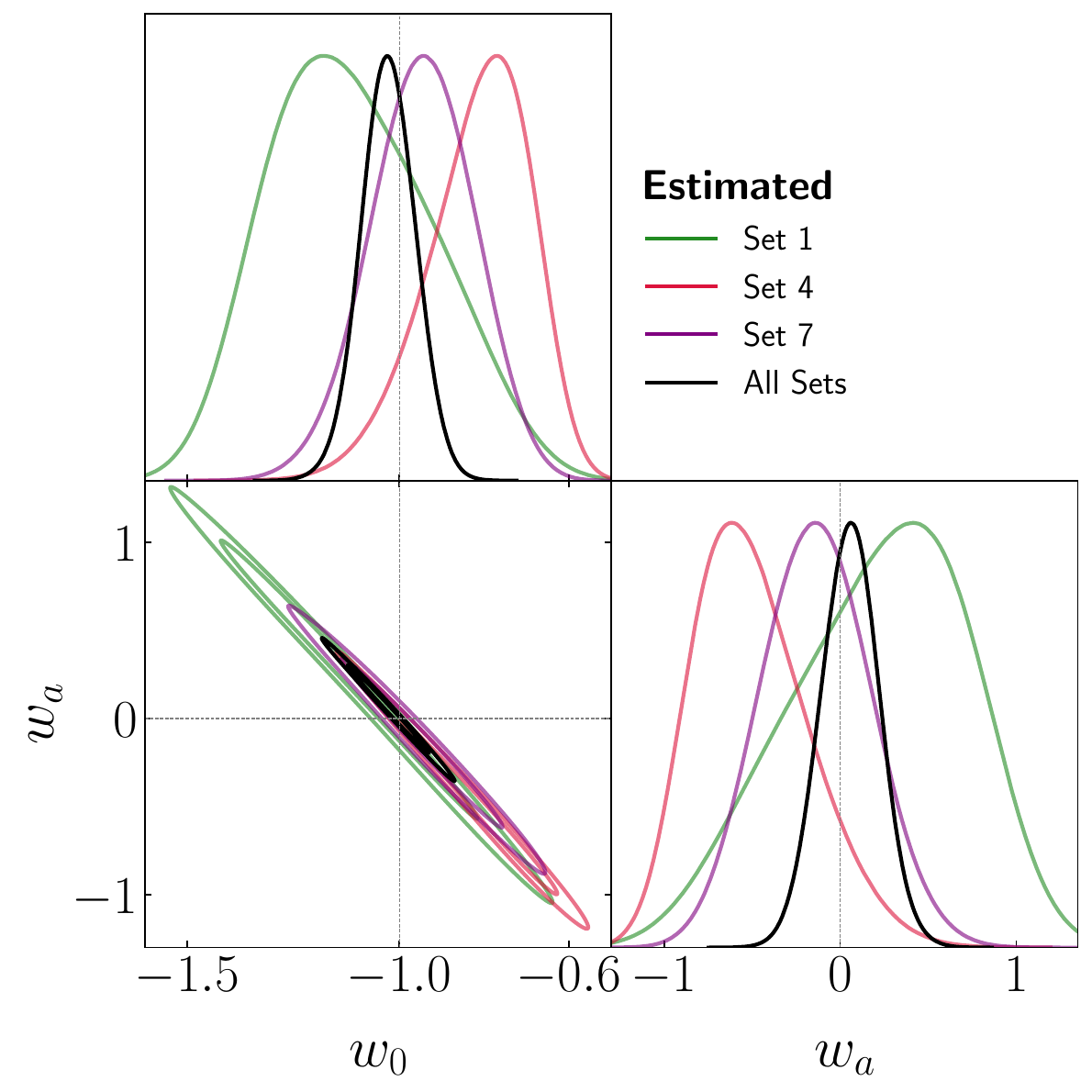} 
    \caption{Posterior distributions of the cosmological parameters $(w_0,w_a)$ for the dynamical dark energy models. The left panel shows the joint parameter estimation from the seven data sets with nineteen cross-power spectra, which are generated from theoretical calculations (``Theory"), Gaussian simulations (``Simulated"), and the realistic simulations with both the foreground and bias mitigated (``Estimated"). In the right panel, we show the individual confidence contours for three selected data sets.}
    \label{fig: triangle_plot_forecast}
\end{figure*}

\section{Inference of dynamical dark energy from future cross-power spectra}
\label{forecast}

As discussed in the previous sections, we can obtain the foreground- and bias-mitigated cross-power spectra which are consistent with the input theory. In this section, we move further to investigate the parameter inference using these cross-power spectra. By running the end-to-end simulations discussed in Section \ref{validations}, we can identify the key issues challenging the requirements of precision 21 cm cosmology. 

The recent DESI results indicate that dark energy may not be a cosmological constant; instead, it could be dynamically evolving~\citep{desi2}. The measurements at radio wavelengths can independently test these theories, and both the auto- and cross-power spectra can be used to infer the cosmological parameters. In a recent work, cosmological inference has been studied using auto-power spectra estimated from HI mock data at the post-ionization epoch~\citep{BINGO_X}. In this work, we investigate whether the cross-power spectra generated from the realistic simulations can still yield unbiased cosmological information. As described in Section \ref{model}, we consider an evolving dark energy with two parameters $w_0$ and $w_a$ and perform a Bayesian analysis with the likelihood function defined as
\begin{equation}
    \chi^2=\displaystyle\sum_{b=1}^{n_b}\Big[\frac{{\hat C}_b^{{\rm HI}, g }-C_b^{{\rm HI}, g }(w_0,w_a)}{\sigma_b}\Big]^2,\label{chi2}
\end{equation}
where $b$ is the band index, $n_b$ is the number of bands, and $\sigma_b$ is calculated from the variance using the amplitude-corrected angular power spectra. In the left panel of Figure \ref{fig: triangle_plot_forecast}, the $w_0\mbox{-}w_a$ contours from three different estimates are consistent with the fiducial values. Moreover, we show the equation-of-state parameter constraints from three individual sets in the right panel. A single set may yield a slightly skewed posterior distribution, which is mainly caused by the non-Gaussian HI reconstruction after foreground removal because of the foreground leakage to the HI component. However, when all sets are combined, the skewness in the one-dimensional likelihood functions of both parameters is almost suppressed. 

To assess the goodness-of-fit of the recovered cross-spectra, we computed $p$-values from the minimum $\chi^2$ obtained at the best-fit equation-of-state parameters, using the band-power errors adopted in this analysis. These $p$-values are interpreted as consistency diagnostics rather than as definitive evidence for or against residual systematics. Therefore, when several data sets or cross-spectra are analyzed, some low $p$-values are expected purely from statistical fluctuations. Furthermore, under the null hypothesis and in the absence of unmodeled effects, the $p$-values are expected to be approximately uniformly distributed. In this work, the seven selected data sets contain nineteen cross-spectra; thus, one would expect approximately one case with a $p<0.05$ by chance alone. Among the analyzed sets, only $\textbf{Set}$1 gives a $p$-value below the 0.05 threshold, with a $p=0.02$. This value is therefore not, by itself, evidence for a residual systematic effect. Moreover, the minimum reduced $\chi^2$ for $\textbf{Set}$1 is consistent with that obtained from the corresponding signal-only Gaussian simulations within the expected scatter of the $\chi^2$ statistic. Together with the absence of a broader excess of low $p$-values across the data sets, this indicates that the recovered spectra are statistically consistent with the model and do not show evidence for a systematic bias.

\section{Conclusions}
\label{conclu}
In this work, we simulate radio sky maps that consist of major Galactic radio contaminants, HI signals, and instrumental effects, and adopt the FastICA method to remove the bright foregrounds from the mock data. We also simulate the narrow-band galaxy surveys and incorporate the expected correlations from theoretical calculations for these two datasets. We cross-correlate the reconstructed 21 cm signals with narrow-band galaxy overdensity fields and estimate the transfer functions resulting from the signal loss in the cross-power spectra. We obtain the foreground- and bias-mitigated HI-galaxy cross-power spectra after applying the corrections with the transfer functions.

We perform the Bayesian analysis with these cross-power spectra and obtain results with no significant biases in the final parameter estimation when combining all selected datasets. This work demonstrates that the narrow-band LSS tracers, such as those achieved from the photometric and spectroscopic galaxy surveys, will be ideal candidates for measuring the HI-LSS cross-power spectra that can be further used for cosmological studies. Also, the analysis formalism developed in this work can be successfully applied to future radio and optical surveys and can yield sensitive measurements of the cross-power spectra, which will be crucial for unraveling the nature of dark energy.

\section*{Acknowledgments}
\begin{acknowledgments}
This work is supported by the USTC's starting grant. We acknowledge the use of the \textsc{Healpix}~\citep{HEALpix}, \textsc{matplotlib}
\citep{matplotlib}, \textsc{numpy} \citep{numpy}, \textsc{scikit-learn} \citep{scikit-learn}, \textsc{CAMB} \citep{camb} and \textsc{NaMaster} \citep{namaster}. 
\end{acknowledgments}

\appendix

\section{The impact of the galaxy full covariance on the transfer function calculations \label{Section: cross-g and TF}}
In the main text, we emphasize that only the diagonal elements of the galaxy covariance are used in the simulations of the correlated fields. We can prove that the inclusion of the off-diagonal elements will not affect the cross-power transfer functions. We denote the HI harmonic coefficients at the frequency channel $v_i$ by $a_{\ell m}^{\rm HI}(v_i)$, and the galaxy harmonic coefficients at bin $b_j$ by $a_{\ell m}^{g}(b_j)$. Although the analysis in this work uses 30 radio-frequency channels and 40 galaxy redshift bins, we keep the notation general and consider $M$-radio channels and $N$-galaxy bins. In this way, we defined the vectors containing all multipoles of the field as ${\boldsymbol a}_{\ell m}^{\rm HI} = [a_{lm}^{\rm HI}(v_1), \dots, a_{lm}^{\rm HI}(v_M)]^{T}$ and ${\boldsymbol a}_{\ell m}^{g} = [a_{lm}^{g}(b_1), \dots, a_{lm}^{g}(b_N)]^{T}$.

To generate correlated realizations from the theoretical angular power spectra, the spectra are used to construct the covariance matrix of the harmonic coefficients at each multipole $\ell$. For a given pair $(\ell,m)$, the harmonic coefficients of all fields can be grouped into a single vector, and the HI and galaxy overdensity components can be defined as
\begin{equation}
{\boldsymbol a}_{\ell m}
=
\begin{bmatrix}
{\boldsymbol a}_{\ell m}^{\rm HI} \\
{\boldsymbol a}_{\ell m}^{g}
\end{bmatrix}
=
\begin{bmatrix}
a_{\ell m}^{\rm HI}(v_1), \dots,
a_{\ell m}^{\rm HI}(v_M), a_{\ell m}^{g}(b_1), \dots,  a_{\ell m}^{g}(b_N)
\end{bmatrix}^{T}.
\end{equation}

From these definitions, one can construct the covariance matrices of the auto- and cross-angular power spectra for the two tracers. If the bandwidth of HI IM channels is sufficiently broad, the correlations among different channels can be neglected, i.e., $C_\ell^{\rm HI}(v_i,v_j)=\delta_{ij}\,C_\ell^{\rm HI}(v_i)$, thus, the HI covariance matrix is essentially diagonal
\begin{equation}
\mathbf{C}_\ell^{\rm HI} 
\equiv
\left\langle
{\boldsymbol a}^{\rm HI}_{\ell m}({\boldsymbol a}^{\rm HI}_{\ell m})^{\dagger}
\right\rangle =
\begin{bmatrix}
    C_\ell^{\rm HI}(v_1) & & \mathbf{0} \\
          & \ddots &  \\
    \mathbf{0} & & C_\ell^{\rm HI}(v_M)
\end{bmatrix}.
\end{equation}
Here, the superscript $\dagger$ denotes the complex conjugate. For the galaxy overdensity field, the binned covariance matrix can be expressed as
\begin{equation}
\mathbf{C}_\ell^{g} 
\equiv
\left\langle
{\boldsymbol a}^{g}_{\ell m}({\boldsymbol a}^g_{\ell m})^{\dagger}
\right\rangle =
\begin{bmatrix}
    C_\ell^{g}(b_1,b_1) & \cdots & C_\ell^{g}(b_1,b_N) \\
    \vdots & \ddots & \vdots \\
    C_\ell^{g}(b_N,b_1) & \cdots & C_\ell^{g}(b_N,b_N)
\end{bmatrix},
\end{equation}
where off-diagonal terms account for correlations between different galaxy bins.

The HI-galaxy cross-covariance matrix is given by
\begin{equation}
\mathbf{C}_\ell^{\rm HI,g} 
\equiv
\left\langle
{\boldsymbol a}^{\rm HI}_{\ell m}({\boldsymbol a}^g_{\ell m})^{\dagger}
\right\rangle =
\begin{bmatrix}
    C_\ell^{\rm HI,g}(v_1,b_1) & \cdots & C_\ell^{\rm HI,g}(v_1,b_N) \\
    \vdots & \ddots & \vdots \\
    C_\ell^{\rm HI,g}(v_M,b_1) & \cdots & C_\ell^{\rm HI,g}(v_M,b_N)
\end{bmatrix}.
\end{equation}

Therefore, the full covariance matrix in harmonic space at each multipole $\ell$ can be written as
\begin{equation}
\boldsymbol{\Sigma}_\ell
\equiv
\left\langle
\mathbf{a}_{\ell m}\mathbf{a}_{\ell m}^{\dagger}
\right\rangle =
\begin{bmatrix}
    \mathbf{C}_\ell^{\rm HI} & \mathbf{C}_\ell^{\rm HI,g} \\
    \left(\mathbf{C}_\ell^{\rm HI,g}\right)^{\dagger} & \mathbf{C}_\ell^{g}
\end{bmatrix}.
\label{eq: (appendix) COV_CL}
\end{equation}

We can perform the Cholesky decomposition of the full covariance matrix $\boldsymbol{\Sigma} = {\boldsymbol L}{\boldsymbol L}^T$ and use the lower triangular matrix ${\boldsymbol L}$ to generate correlated fields via
\begin{eqnarray}
    {\boldsymbol a}_{\ell m} = {\boldsymbol L}_{\ell m} {\boldsymbol \xi}_{\ell m},
\label{eq: (appendix) alm_cholesky}    
\end{eqnarray}
where the ${\boldsymbol \xi}_{\ell m} = [(\xi_{1})_{\ell m}, \dots, (\xi_{N+M})_{\ell m}]^{T}$ is a complex random (M+N)-vector, whose terms are orthogonal to each other. Also, the lower triangular matrix ${\boldsymbol L}$ is constructed from three sub-matrices as 
\begin{eqnarray}
\begin{pmatrix} {\boldsymbol a}_{\ell m}^{\rm HI} \\ {\boldsymbol a}_{\ell m}^g \end{pmatrix} = 
\begin{pmatrix} ({\boldsymbol L_A})_{\ell m} & \mathbf{0} \\ ({\boldsymbol \Delta})_{\ell m} & ({\boldsymbol L_B})_{\ell m} \end{pmatrix}
\begin{pmatrix} {\boldsymbol \xi_{lm}^{(1)}} \\ {\boldsymbol \xi_{lm}^{(2)}}\end{pmatrix},
\label{eq: (appendix) alm_L_xi__matricial}
\end{eqnarray}
where we introduce matrices ${\boldsymbol L_A}$, ${\boldsymbol L_B}$, and ${\boldsymbol \Delta}$. Similarly, we can split the vector ${\boldsymbol \xi}$ into two vectors, ${\boldsymbol \xi^{(1)}}$ and ${\boldsymbol \xi^{(2)}}$ which are defined as 
\begin{align}
    {\boldsymbol \xi^i} &\equiv {\boldsymbol \xi^{i(1)}}, \quad i \in [1, M] \\
    {\boldsymbol \xi^{M+i}} &\equiv {\boldsymbol \xi^{i(2)}}, \quad i \in [M+1, M+N].
\end{align}
Here, each vector satisfies the relation $\langle \xi^{i(a),\ast}_{\ell m}, \xi^{j(b)}_{\ell' m'}\rangle=\delta_{ij}\delta_{ab}\delta_{\ell\ell'}\delta_{m m'}$.

With these definitions, we can re-derive the full covariance matrix in terms of the ${\boldsymbol L}$ elements as
\begin{align}
    {\boldsymbol \Sigma} &= \begin{pmatrix}
                {\boldsymbol L_A} {\boldsymbol L_A}^T & ({\boldsymbol L_A} {\boldsymbol \Delta}^T) \\
                ({\boldsymbol L_A} {\boldsymbol \Delta}^T)^T & {\boldsymbol \Delta}{\boldsymbol \Delta}^T + {\boldsymbol L_B}{\boldsymbol L_B}^T
             \end{pmatrix}.
\label{eq: (appendix) COV_LL}
\end{align}

Comparing the Eqs. (\ref{eq: (appendix) COV_CL}) with (\ref{eq: (appendix) COV_LL}), we obtain the relations: ${\boldsymbol C}_{\ell}^{\rm HI} = {\boldsymbol L_A} {\boldsymbol L_A}^T$, ${\boldsymbol C}_{\ell}^{\rm HI,g} = {\boldsymbol L_A} {\boldsymbol \Delta}^T$, and ${\boldsymbol C}_{\ell}^g = {\boldsymbol \Delta}{\boldsymbol \Delta}^T + {\boldsymbol L_B} {\boldsymbol L_B}^T$. For given covariance matrices ${\boldsymbol C}_{\ell}^{\rm HI}$ and ${\boldsymbol C}_{\ell}^{\rm HI,g}$, the two sub-matrices ${\boldsymbol L_A}$ and ${\boldsymbol \Delta}$ are thus fixed. Therefore, the off-diagonal elements of the galaxy full covariance matrix will be absorbed in the ${\boldsymbol L_B}$ matrix, which does neither affect the ${\boldsymbol L_A}$ matrix nor the ${\boldsymbol \Delta}$ matrix. 

\bibliography{biblio}{}
\bibliographystyle{aasjournal}
\end{document}